\documentclass[pra,twocolumn,showpacs]{revtex4}

\usepackage{graphics}       
\usepackage{bm}         
\usepackage{amsmath}        
\usepackage{latexsym}
\begin{document}
\title{Distinguishability Measures and Ensemble Orderings}
\author{Anthony Chefles}
\affiliation{Department of Physical Sciences, University of
Hertfordshire,
       Hatfield AL10 9AB, Hertfordshire, UK}

\begin{abstract}
\vspace{0.5cm}.
\\It is shown that different distinguishability measures impose
different orderings on ensembles of $N$ pure quantum states.  This
is demonstrated using ensembles of equally-probable, linearly
independent, symmetrical pure states, with the maximum
probabilities of correct hypothesis testing and unambiguous state
discrimination being the distinguishability measures.  This
finding implies that there is no absolute scale for comparing the
distinguishability of any two ensembles of $N$ quantum states, and
that
 distinguishability comparison is necessarily relative to a particular discrimination strategy.

\end{abstract}
 \maketitle

\section{Introduction}
\renewcommand{\theequation}{1.\arabic{equation}}
\setcounter{equation}{0}

The use of quantum channels to send classical information has many
advantages over the use of classical channels.  One of these is
the fact that it enables one to establish cryptographic keys in a
way that is provably secure, a feat which has never been achieved
classically. Developments such as this have led to renewed
attention being given to the problem of distinguishing between
quantum states\cite{Review}. In state discrimination, we are faced
with the following situation: a quantum system is prepared in one
of $N$ states.  For the sake of simplicity, we will take these to
be pure states $|{\psi}_{j}{\rangle}$, where $j=1,{\ldots},N$. The
{\em a priori} probability of the state of the system being
$|{\psi}_{j}{\rangle}$ is $p_{j}$.  We would like to determine
which state has been prepared.  Unless the states are orthogonal
we cannot determine the state perfectly.  We are then faced with
the problem of devising a strategy which discriminates between the
$N$ potential states as well as possible. This will involve some,
possibly generalised, quantum measurement. This measurement should
be optimised and the resulting figure of merit, which is typically
a probability, can be regarded as a measure of the
distinguishability of the states with these {\em a priori}
probabilities.  A set of quantum states $|{\psi}_{j}{\rangle}$
considered together with their {\em a priori} probabilities
$p_{j}$ forms an ensemble ${\cal E}={\cal
E}(p_{j},|{\psi}_{j}{\rangle})$. So, distinguishability measures
will refer to ensembles.

We shall denote a generic distinguishability measure by $D[{\cal E}]$. Several distinct measures are in common use for quantifying the
distinguishability of states\cite{Review}.  The question we address in this paper is: do all distinguishability measures impose the
same ordering on ensembles of $N$ quantum states? That is, suppose that we have two ensembles, ${\cal E}_{1}$ and ${\cal E}_{2}$, and
two distinguishability measures, $D_{1}$ and $D_{2}$. Then, if $D_{1}[{\cal E}_{1}]>D_{1}[{\cal E}_{2}]$, is it the case that
$D_{2}[{\cal E}_{1}]{\geq}D_{2}[{\cal E}_{2}]$? We shall see that this is not necessarily so.

The distinguishability measures we choose are the maximum
probabilities of correct hypothesis testing and unambiguous state
discrimination.  For the sake of simplicity, we restrict our
attention to ensembles of states which have equal {\em a priori}
probabilities.  We show that for two equally-probable pure states,
this effect cannot be demonstrated for these distinguishability
measures, which leads us to consider ensembles of $N>2$ states. We
restrict our attention to linearly independent ensembles, since
this is a requirement for unambiguous state
discrimination\cite{Chefles1}. The states in the chosen ensembles
are taken to form symmetrical sets. This is done to take advantage
of the fact that for such ensembles, our distinguishability
measures can be calculated analytically for equal {\em a priori}
probabilities. The effect we have described is shown to occur for
$N=3$. We conclude with a discussion of the connection between
this effect and a related one recently discovered by Jozsa and
Schlienz\cite{JS}.

\section{Distinguishability Measures Imposing Different Ensemble Orderings}
\renewcommand{\theequation}{2.\arabic{equation}}
\setcounter{equation}{0}

The distinguishability measures we choose are the maximum
probabilities of correct hypothesis testing and unambiguous state
discrimination.  In hypothesis testing among $N$ states in the
ensemble ${\cal E}(p_{j},|{\psi}_{j}{\rangle})$, we consider an
$N$ outcome generalised measurement, in which the $j$th outcome is
associated with the POVM element $E_{j}$.  Our hypothesis is that
the outcome of the measurement corresponds exactly to the state.
If the state is $|{\psi}_{j}{\rangle}$ and outcome $j$ is
obtained, then the hypothesis is correct.  If however, outcome
$j'{\neq}j$ is obtained, then the hypothesis will be incorrect.
The maximum probability $P_{HYP}$ that our hypothesis is correct
is\cite{Helstrom}
\begin{equation}
P_{HYP}({\cal E})={\max_{
\{E_{j}\}}}\sum_{j}p_{j}{\langle}{\psi}_{j}|E_{j}|{\psi}_{j}{\rangle}
\end{equation}
where the maximisation is carried out over all sets of $N$
positive operators $E_{j}$ such that $\sum_{j}E_{j}=1$.  We shall
use the maximum probability of correct hypothesis testing,
$P_{HYP}({\cal E})$, as a measure of the distinguishability of the
ensemble ${\cal E}$.

In unambiguous state discrimination, there are only two possible
outcomes for the state $|{\psi}_{j}{\rangle}$: outcome $j$ and a
further inconclusive result `?'.  There are no errors. Unlike
hypothesis testing, unambiguous discrimination is only possible
for linearly independent sets\cite{Chefles1} and it is to such
sets that we will restrict our attention.  If
$P_{USD}(|{\psi}_{j}{\rangle})$ is the probability that, given the
initial state was $|{\psi}_{j}{\rangle}$, we obtain a conclusive
identification rather than an inconclusive result, then the
maximum probability of unambiguous state discrimination is

\begin{equation}
P_{USD}({\cal E})={\max_{
\{P_{USD}(|{\psi}_{j}{\rangle})\}}}\sum_{j}p_{j}P_{USD}(|{\psi}_{j}{\rangle}),
\end{equation}
where the extremisation with respect to the
$P_{USD}(|{\psi}_{j}{\rangle})$ is discussed by Duan and
Guo\cite{Duanguo}.  The distinguishability measure we will use for
this strategy will be the maximum probability $P_{USD}({\cal E})$
of unambiguous determination of the state.

Here, we will show that there exist ensemble pairs for which
\begin{eqnarray}
P_{USD}({\cal E}_{2})&<&P_{USD}({\cal E}_{1}), \\ P_{HYP}({\cal
E}_{2})&>&P_{HYP}({\cal E}_{1}).
\end{eqnarray}
The effect we wish to demonstrate does not occur for an ensemble
of just two equally-probable pure states.  This can be seen by
examining the values of $P_{HYP}$ and $P_{USD}$ for a pair of pure
states, $|{\psi}_{1}{\rangle}$ and $|{\psi}_{2}{\rangle}$.  For
later convenience, we will give expressions for these with
arbitrary {\em a priori} probabilities $p_{1}$ and $p_{2}$. The
former is given by Helstrom's bound\cite{Helstrom}:
\begin{equation}
P_{HYP}=\frac{1}{2}\left(
1+\sqrt{1-(1-{\Delta}^{2})|{\langle}{\psi}_{1}|{\psi}_{2}{\rangle}|^{2}}\right),
\end{equation}
where ${\Delta}=|p_{1}-p_{2}|$.  Also, the maximum value of
$P_{USD}$ for a pair of pure states is given by the Jaeger-Shimony
bound\cite{JShimony}
\begin{equation}
P_{USD}=\left\{
\begin{array}{cc}
1-\sqrt{1-{\Delta}^{2}}|{\langle}{\psi}_{1}|{\psi}_{2}{\rangle}|&: \sqrt{\frac{1-{\Delta}}{1+{\Delta}}}{\geq}|{\langle}{\psi}_{1}|{\psi}_{2}{\rangle}|\; \\
\frac{1}{2}(1+{\Delta})(1-|{\langle}{\psi}_{1}|{\psi}_{2}{\rangle}|^{2})
&:\sqrt{\frac{1-{\Delta}}{1+{\Delta}}}{\leq}|{\langle}{\psi}_{1}|{\psi}_{2}{\rangle}|\;.
\end{array}
\right.
\end{equation}
When the a priori probabilities are equal, ${\Delta}=0$ and
$P_{HYP}=(1/2)(1+\sqrt{1-(1-(1-P_{USD})^{2}})$. From this, one can
show that $P_{HYP}$ is an increasing function of $P_{USD}$,
implying that inequalities (2.3) and (2.4) can never be
simultaneously satisfied.

To find ensembles of equally-probable pure states for which both
(2.3) and (2.4) are true, we have to consider at least three
states. We focus on ensembles of equally-probable, linearly
independent, symmetrical states, since the maximum probabilities
for unambiguous discrimination and correct hypothesis testing can
be calculated explicitly for these. We will demonstrate the
existence of ensemble pairs ${\cal E}_{1}$ and ${\cal E}_{2}$
which satisfy inequalities (2.3) and (2.4), in the following way.
Firstly, we will consider {\em all} sets of $N$ linearly
independent, symmetric states with equal {\em a priori}
probabilities which have the same, arbitrary but fixed, value of
$P_{USD}$.  Over this set, we will find the extremal values of
$P_{HYP}$.  Using this information, we choose ensembles pairs
which satisfy inequality (2.3), but where ${\cal E}_{1}$ and
${\cal E}_{2}$ have respectively the minimum and maximum values of
$P_{HYP}$ for their corresponding values of $P_{USD}$. We will
find that, for $N=3$, inequality (2.4) is satisfied for a large
range of parameters.

The $N$ pure states $|{\psi}_{j}{\rangle}$, where
$j=0,{\ldots},N-1$, are linearly independent and symmetric if and
only if they can be written as
\begin{equation}
|{\psi}_{j}{\rangle}=\sum_{r=0}^{N-1}c_{r}e^{\frac{2{\pi}ijr}{N}}|x_{r}{\rangle},
\end{equation}
for some $N$ orthonormal states $|x_{r}{\rangle}$ and non-zero
complex coefficients $c_{r}$ satisfying $\sum_{r}|c_{r}|^{2}=1$.
Notice that the phase of $c_{r}$ may be absorbed by
$|x_{r}{\rangle}$, which implies that we may, without loss of
generality, take $c_{r}$ to be real and positive, which we shall.
We will take all states to have equal {\em a priori} probabilities
$p_{j}=1/N$. The maximum unambiguous discrimination probability
for these states is\cite{Mesymmetric}
\begin{equation}
P_{USD}=N{\times}\min_{r}c_{r}^{2}.
\end{equation}
The optimum hypothesis testing strategy uses the so-called
`square-root' measurement\cite{Helstrom}.  Define the operator
\begin{equation}
{\Phi}=\sum_{j}|{\psi}_{j}{\rangle}{\langle}{\psi}_{j}|.
\end{equation}
and the states
\begin{equation}
|{\omega}_{j}{\rangle}={\Phi}^{-1/2}|{\psi}_{j}{\rangle}.
\end{equation}
One can quite easily show that the operators
$E_{r}=|{\omega}_{r}{\rangle}{\langle}{\omega}_{r}|$ form a POVM
(i.e., that $E_{r}{\geq}0$ and $\sum_{r}E_{r}=1$.) This POVM is
the optimal hypothesis testing strategy, and we find that
\begin{equation}
P_{HYP}=\frac{1}{N}\left(\sum_{r}c_{r}\right)^{2}.
\end{equation}
The square-root measurement optimally discriminates between any
set of equally-probable, symmetric states, even if they are not
linearly independent.  This measurement has recently been carried
out\cite{Expt1,Expt2} for symmetrical optical polarisation states.
Applications of this measurement to quantum key distribution are
discussed in\cite{Crypt}.

We now calculate the global extrema of $P_{HYP}$ for a fixed value
of $P_{USD}$. Our aim is to fix the smallest of the $c_{r}$, which
is equivalent to fixing $P_{USD}$, and vary the remaining
coefficients to obtain the extremal values of $P_{HYP}$. We may
let ${\min_r}c_{r}=c_{0}$. This allows us to write
\begin{equation}
c_{r}=c_{0}+{\cos}^{2}{\theta}_{r},
\end{equation}
for $r=1,{\ldots},N-1$ and some angles ${\theta}_{j}$.  We will
now extremise $P_{HYP}$ with respect to these angles using the
method of Lagrange multipliers, in order to take into account the
normalisation constraint.  This method will yield the local
extrema, over which we will subsequently optimise to find the
global extrema. Let $G=(\sum_{r=0}^{N-1}c_{r}^{2})-1$. The
constrained, local extrema of $P_{HYP}$ occur where
\begin{equation}
\frac{\partial P_{HYP}}{\partial
\theta_{r}}={\lambda}\frac{\partial G}{\partial \theta_{r}}
\end{equation}
and ${\lambda}$ is our Lagrange multiplier.  Inserting Eq. (2.12)
into the definitions of $P_{HYP}$ and $G$, we find that this
becomes
\begin{equation}
{\sin}{\theta}_{r}{\cos}{\theta}_{r}\left[c_{0}+\frac{1}{N}\sum_{r'=1}^{N-1}{\cos}^{2}{\theta}_{r'}-{\lambda}(c_{0}+{\cos}^{2}{\theta}_{r})\right]=0.
\end{equation}
It is a simple matter to show that the normalisation requirement
will be violated if ${\sin}{\theta}_{r}=0$ for any $r$.  So, for
each $r$, either ${\cos}{\theta}_{r}=0$, in which case
$c_{r}=c_{0}$, or the term in brackets is zero.  To proceed, we
will partition the $c_{r}$ into two sets, where each set
corresponds to one of these two possibilities.  Let there be
$N_{0}$ coefficients, including $c_{0}$, which are equal to
$c_{0}$.  For the remaining $N-N_{0}$ coefficients, the term in
brackets is zero. All of the coefficients in this latter set must
also be equal, as can be seen from the fact that the vanishing of
the bracket term implies that the corresponding
${\cos}^{2}{\theta}_{r}$ are equal. The value that they have is
easily deduced from the fact that the other $N_{0}$ coefficients
are equal to $c_{0}$, and normalisation.  We find they have the
value
\begin{equation}
c_{r}=\sqrt{\frac{1-N_{0}c_{0}^{2}}{N-N_{0}}}.
\end{equation}
We can now calculate the local extrema of $P_{HYP}$.  Evaluating
Eq. (2.11) where $N_{0}$ coefficients are equal to $c_{0}$, with
the remaining $N-N_{0}$ coefficients being given by Eq. (2.15),
and making use of the fact that $c_{0}=\sqrt{P_{USD}/N}$, we find
that the local extrema of $P_{HYP}$ for fixed $P_{USD}$ are
\begin{equation}
P_{HYP}=\frac{1}{N^{2}}\left(N_{0}\sqrt{P_{USD}}+\sqrt{(N-N_{0})(N-N_{0}P_{USD})}\right)^{2}
\end{equation}
Finding the global extrema of $P_{HYP}$ amounts to extremisation
of this expression with respect to $N_{0}$. One method of doing
this is to treat $N_{0}$ as a continuous parameter in the interval
$[1,N-1]$ and differentiate Eq. (2.16) with respect to it. After
some algebra, we find that
${\partial}P_{HYP}/{\partial}N_{0}{\leq}0$, for $P_{HYP}$ given by
Eq. (2.16).  This implies that the global maximum and minimum of
$P_{HYP}$ occur at the minimum and maximum values of $N_{0}$
respectively.  Thus, the maximum value of $P_{HYP}$ occurs at
$N_{0}=1$, which gives us the tight upper bound

\begin{equation}
P_{HYP}{\leq}\frac{1}{N^{2}}\left(\sqrt{P_{USD}}+\sqrt{(N-1)(N-P_{USD})}\right)^{2}.
\end{equation}
For $P_{USD}=1$, we see that the maximum of $P_{HYP}$ is also
equal to 1. However, when $P_{USD}=0$, the maximum of $P_{HYP}$ is
$(N-1)/N$.

\begin{figure}
\scalebox{0.9}[0.9]{\includegraphics{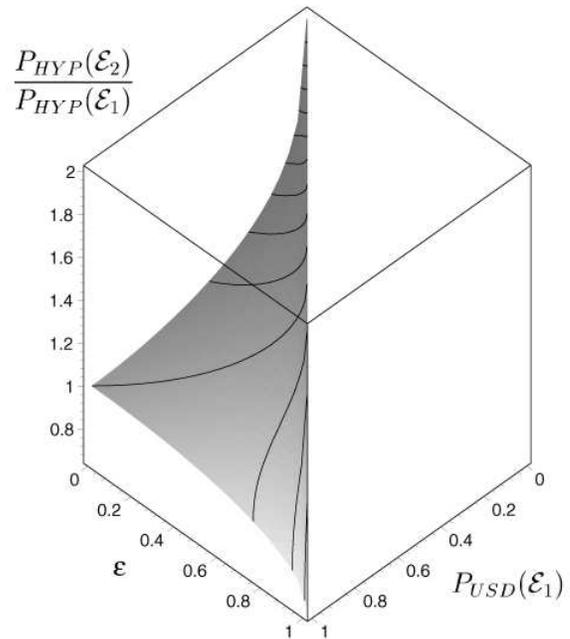}} \vspace{0.5in}
\caption{Contour plot of the ratio of the maximum probabilities of
correct hypothesis testing for ensembles ${\cal E}_{1}$ and ${\cal
E}_{2}$ of three linearly independent, equally-probable, symmetric
states, versus the maximum unambiguous discrimination probability
$P_{USD}({\cal E}_{1})$ of ${\cal E}_{1}$ and the parameter
${\epsilon}=P_{USD}({\cal E}_{1})-P_{USD}({\cal E}_{2})$.  This
parameter is positive which implies that, except at the origin,
inequality (2.3) is satisfied.  Inequality (2.4) is satisfied
whenever the ratio $P_{HYP}({\cal E}_{2})/P_{HYP}({\cal
E}_{1})>1$, which, on this plot, corresponds to points lying above
the fourth contour from the bottom.}

\end{figure}

We also see that the minimum value of $P_{HYP}$ occurs at
$N_{0}=N-1$, which gives us the tight lower bound
\begin{equation}
P_{HYP}{\geq}\frac{1}{N^{2}}\left((N-1)\sqrt{P_{USD}}+\sqrt{N-(N-1)P_{USD}}\right)^{2}.
\end{equation}
When $P_{USD}=1$, the minimum of $P_{HYP}$ is also 1, as we would
expect.  However, if $P_{USD}=0$, then the minimum of $P_{HYP}$ is
$1/N$, which corresponds to a random guess of which of the $N$
possible states the system has been prepared in.  In fact, it is
quite easily shown that in this case, all $|{\psi}_{j}{\rangle}$
are, up to a phase, equal to $|x_{0}{\rangle}$.  Here, the states
are linearly dependent and this case is of no interest to us in
the present context.

 Let us now use the bounds in (2.17)
and (2.18) to demonstrate the existence of ensembles for which
inequalities (2.3) and (2.4) hold.  For the sake of simplicity, we
will consider only ensembles for which $N=3$.  We will choose
${\cal E}_{2}$ to be an ensemble which saturates the bound in
(2.17) and ${\cal E}_{1}$ to be one which saturates the bound in
(2.18), yet where $P_{USD}({\cal E}_{2})=P_{USD}({\cal
E}_{1})-{\epsilon}$, for some positive, non-zero parameter
${\epsilon}$.  This guarantees that inequality (2.3) is satisfied.
Figure 1 depicts $P_{HYP}({\cal E}_{2})/P_{HYP}({\cal E}_{1})$ as
a function of $P_{USD}({\cal E}_{1})$ and ${\epsilon}$, where
$0{\leq}{\epsilon}{\leq}P_{USD}({\cal E}_{1})$ and $N=3$.
Satisfaction of the inequality (2.4) occurs for parameters where
$P_{HYP}({\cal E}_{2})/P_{HYP}({\cal E}_{1})>1$, and a large range
of such parameters in clearly visible in the figure.

For example, let $P_{USD}({\cal E}_{1})=0.5$ and $P_{USD}({\cal
E}_{2})=0.4$.  This gives ${\epsilon}=0.1$ and inequality (2.3) is
satisfied.  Evaluating the maximum correct hypothesis testing
probabilities, we find that $P_{HYP}({\cal E}_{1})=8/9{\sim}0.888$
and $P_{HYP}({\cal E}_{2}){\sim}0.943>P_{HYP}({\cal E}_{1})$ which
satisfies inequality (2.4).

\section{Distinguishability and Information}
\renewcommand{\theequation}{3.\arabic{equation}}
\setcounter{equation}{0}

We have seen that different distinguishability measures impose
different orderings on ensembles of pure quantum states.  We
considered linearly independent states and demonstrated this
effect using the maximum probabilities of correct hypothesis
testing and unambiguous state discrimination as distiguishability
measures.

This result has important implications for any situation where we
wish to transmit nonorthogonal states to send classical
information.  For example, in quantum key distribution, we may
have a choice of sending states prepared in either ensemble ${\cal
E}_{1}$ or ensemble ${\cal E}_{2}$.  The distinguishability of
these ensembles may determine the potential information available
to an eavesdropper and also the rate at which key bits can be
generated.  It would therefore be desirable to able to compare the
distinguishability of both ensembles.  What the results in the
preceding show is that this cannot be done in any absolute sense;
we must know in advance which distinguishability measure is being
used. However, we may not know which measure would apply to a
potential eavesdropper since we may not know in advance which
detection strategy they would employ.

To put the results of the preceding section in context, it is
helpful to consider a related finding recently made by Jozsa and
Schlienz\cite{JS}.  They considered two ensembles of pure states,
${\cal E}_{1}$ and ${\cal E}_{2}$, with all states in each
ensemble having equal {\em a priori} probabilities. They used the
pairwise overlaps of pairs of states as a means of quantifying the
distinguishability of each ensemble, and the von Neumann entropy
$S_{i}$ of the density operator of ensemble ${\cal E}_{i}$ as a
means of quantifying information content. They showed that there
exist ensemble pairs in which the pairwise overlaps, and hence the
distiguishability of ${\cal E}_{2}$ are less than those of ${\cal
E}_{1}$, yet where $S_{2}>S_{1}$, implying that the information
content of ${\cal E}_{2}$ is greater than that of ${\cal E}_{1}$.
This is somewhat counterintuitive, as information and
distinguishability are often regarded as interchangeable concepts.

This finding suggests that there are distinctions to be made
between information and distinguishability.  That they are not
entirely interchangeable is perhaps suggested by the fact that
they arise in different contexts.  In attempting to distinguish
between a set of states, we are given only one copy, and must make
optimum use of it.  In information transmission however, we
perform a collective measurement on a large number of states drawn
from the same ensemble.  The length of the strings and the subset
chosen from the set of all possible strings is such that these
strings are highly distinguishable.  Indeed, they are chosen in
such a way that the probability of failing to distinguish between
the states perfectly can be made arbitrarily small.   The
information content is the number of bits of classical information
that can be transmitted per signal in each string, where the
signals obey the statistical constraints of the ensemble.  In
information transmission, we select strings which are sufficiently
long so that we can neglect the distinguishability issue in one
way. However, it is possible that this issue does return in some
form to dictate the information content of a single signal in ways
that are not currently well-understood.

A conclusive distinction, which does not apply to situations in
which all states have the same {\em a priori} probabilities, is
that information content and distinguishability can vary in
opposite ways when the {\em a priori} probabilities of the states
are altered.  To illustrate this, consider two ensembles, ${\cal
E}_{1}$ and ${\cal E}_{2}$, where both consist of the pure states
$|{\psi}_{1}{\rangle}$ and $|{\psi}_{2}{\rangle}$.  In ${\cal
E}_{1}$, both states have equal {\em a priori} probabilities equal
to 1/2.  However, in ${\cal E}_{2}$, they have unequal {\em a
priori} probabilities $p_{1}$ and $p_{2}$.  In state
discrimination, our goal is to determine the state.  If one state
is more probable than the other, then we can take advantage of
this fact to tailor our measurement to weight it in favour of the
state which has higher {\em a priori} probability, so that, on
average, we will be able to improve our ability to determine the
state.  It follows that ${\cal E}_{2}$ ought to have higher
distinguishability than ${\cal E}_{1}$.  This is made explicit if
we look back at either the Helstrom bound (2.5) or the
Jaeger-Shimony bound (2.6).  These both increase as
${\Delta}=|p_{1}-p_{2}|$ increases, confirming this expectation.

However, if we wish to use states in ensemble ${\cal E}_{2}$ to
send messages, then the constraint that one state must always be
more probable than the other implies that, to regain our freedom
in the message we might choose to send, we must send more signal
states per message as ${\Delta}$ increases.  Quantitatively, for
${\cal E}_{2}$, the information content is given by the von
Neumann entropy of the ensemble density operator, which is equal
to the binary entropy function
\begin{equation}
S_{2}=-x{\log}_{2}x-(1-x){\log}_{2}(1-x)
\end{equation}
where
\begin{equation}
x=\frac{1}{2}\left(1+\sqrt{1-(1-{\Delta}^{2})(1-|{\langle}{\psi}_{1}|{\psi}_{2}{\rangle}|^{2})}\right).
\end{equation}
As ${\Delta}$ increases, $x$ also increases which implies that, as
can easily be shown, $S_{2}$ decreases.  The extreme situation is
where one of the {\em a priori} probabilities, say $p_{1}$, is
equal to 1, in which case $p_{2}=0$.  Here, we know in advance
what the state is, and so we can always determine it perfectly. We
may say that this ensemble has perfect distinguishability.
However, the sender has no freedom in which state to send and the
state is always entirely predictable.  For this reason, it is
impossible to transmit any information using this ensemble, for
which $S_{2}$ is easily seen to be zero.

As the above argument demonstrates, a clear distinction between
information content and distinguishability emerges if variation of
the {\em a priori} probabilities of the states is permitted.
However, it is not entirely clear why they should be distinct
concepts when the {\em a priori} probabilities of all states are
equal. Consequently, for ensembles with equal a priori
probabilities, the observation that information and
distinguishability impose different orderings remains to be
properly understood.

The results in this paper suggest a potential explanation of this
phenomenon.  Given that it is not clear why information content
should not be regarded as a distinguishability measure when we
restrict our attention to ensembles of states with equal {\em a
priori} probabilities, let us assume that it is in fact a suitable
distinguishability measure under these circumstances.  The
Jozsa-Schlienz effect could then be explained as a demonstration
of the fact that, even for ensembles of equally-probable states,
different distinguishability measures impose different ensemble
orderings, which is what we have shown in this paper, using
quantities which are indisputably suitable distinguishability
measures.

It remains to be understood why different distinguishability
measures impose different orderings.  Perhaps, for two
distinguishability measures to be non-trivially distinct from one
another, they must be sensitive to different aspects of ensembles
of states, which can vary, with some degree of independence, from
one ensemble to another.  This seems reminiscent of the discovery
recently made by Virmani and Plenio\cite{Virmani} that different,
good measures of the entanglement of mixed states must order these
states differently.
\section{discussion}
We have shown in this paper that two of the most common measures
of the distinguishability of states, the maximum probabilities of
correct hypothesis testing and unambiguous state discrimination,
are essentially incompatible with each other.  By this, we mean
that they do not impose the same ordering on ensembles of pure
states.  It is possible to have one ensemble, ${\cal E}_{1}$,
which is more distinguishable than another ensemble ${\cal E}_{2}$
when the states must be distinguished unambiguously, yet where
${\cal E}_{2}$ is more distinguishable than ${\cal E}_{1}$ if we
wish to identify the state by minimum error hypothesis testing. In
general, we cannot, in any absolute sense, characterise one
ensemble of states as being more distinguishable than another.
Distinguishability comparison must necessarily refer to a
particular discrimination strategy.

As with any interesting phenomenon, it is important to determine
the conditions under which this effect can be demonstrated
optimally.  Here, we are faced with the fact that, for generic
ensembles of states, the optimisation problems which must be
solved to obtain values of distinguishability measures are
difficult to tackle analytically.  At the time of writing, the
only ensembles for $N>2$ for which both of our chosen
distinguishability measures can be calculated analytically are
ensembles of equally-probable, linearly independent, symmetric
pure states. For such ensembles, inequalities (2.16) and (2.17)
give the extremal values of the correct hypothesis testing
probability for an fixed value of the maximum  probability of
unambiguous state discrimination, and so the optimal conditions
for demonstrating this effect for these ensembles are given by
these relations.  We also explored in some detail the case of
$N=3$.

The effect we have demonstrated is somewhat reminiscent of a
related one recently discovered by Jozsa and Schlienz.  These
authors showed that one can construct an ensemble pair, ${\cal
E}_{1}, {\cal E}_{2}$, where the distinguishability of ${\cal
E}_{1}$, as measured by the pairwise overlaps of states, is
greater than that of ${\cal E}_{2}$, yet where the information
content of ${\cal E}_{2}$, which is quantified by the von Neumann
entropy, is greater than that of ${\cal E}_{1}$.

This is an important finding, not least because the concepts of
distinguishability and information content are sometimes used
interchangeably.  We described how essential differences between
these concepts do arise when, in moving from one ensemble to
another, the a priori probabilities of the states are changed.
However, if the a priori probabilities of all states in both
ensembles are the same, it is by no means clear that, despite the
contrast between the asymptotic nature of information and the `one
shot' nature of distinguishability, information measures cannot
also serve as measures of distinguishability.

Jozsa and Schlienz demonstrated their effect using such ensembles.
What the results in this paper suggest is a possible
interpretation of the Jozsa-Schlienz effect. They showed that, for
ensembles of equally-probable pure states, information and a
particular distinguishability measure suffer an ordering
incompatibility problem.  Does this imply that for such ensembles,
information content is not a suitable distinguishability measure?
The results in this paper suggest that this is not necessarily the
case, since different distinguishability measures suffer analogous
ordering problems.  It could then be the case that, for the
ensembles considered by Jozsa and Schlienz, information content is
in fact a suitable distinguishability measure and that the
ordering incompatibility they discovered is a consequence of the
fact that such problems arise with regard to generic
distinguishbility measures.

\section*{Acknowledgements}
The author would like to thank Masahide Sasaki and Martin B.
Plenio for interesting discussions. This work was supported by the
UK Engineering and Physical Sciences Research Council.


\begin{thebibliography}{99}
\bibitem{Review}
A. Chefles, {\it Contemp. Phys.} {\bf 41} 401 (2000).
\bibitem{Chefles1}
A. Chefles, {\em Phys. Lett. A} {\bf 239} 339 (1998).
\bibitem{JS}
R. Jozsa and J. Schlienz, {\it Phys. Rev. A.} {\bf 62} 012301
(2000).
\bibitem{Helstrom}
C. W. Helstrom, {\em Quantum Detection and Estimation Theory},
(Academic Press, New York, 1976).
\bibitem{Duanguo}
L.- M. Duan and G.- C. Guo, {\em Phys. Rev. Lett.} {\bf 80} 4999
(1998).
\bibitem{JShimony}
G. Jaeger and A. Shimony, {\em Phys. Lett. A} {\bf 197} 83 (1995).
\bibitem{Mesymmetric}
A. Chefles and S. M. Barnett, {\em Phys. Lett. A} {\bf 250} 223
(1998).
\bibitem{Expt1}
R. B. M. Clarke, V. M. Kendon, A. Chefles, S. M. Barnett, E. Riis
and M. Sasaki, {\em Phys. Rev. A} {\bf 64} 012303 (2001).
\bibitem{Expt2}
J. Mizuno, M. Fujiwara, M. Akiba, T. Kawanishi, S. M. Barnett and
M. Sasaki, `Optimum detection for extracting maximum information
from symmetric qubit sets', LANL eprint quant-ph/0106164.
\bibitem{Crypt}
S. J. D. Phoenix, S. M. Barnett and A. Chefles, {\em J. Mod. Opt.}
{\bf 47} 507 (2000).
\bibitem{Virmani}
S. Virmani and M. B. Plenio, {\em Phys. Lett. A} {\bf 268} 31
(2000).

\end{thebibliography}
\end{document}